# Boosting Photodetection via Plasmonic Coupling in Quasi-2D Mixed-n Ruddlesden-Popper Perovskite Nanostripes

Brindhu Malani S,*,a Eugen Klein,a Ronja Maria Piehler,a Rostyslav Lesyuk ab and Christian Klinke*,acd

a *Institute of Physics, University of Rostock, Albert-Einstein-Straße 23, 18059 Rostock, Germany*

b *Pidstryhach Institute for Applied Problems of Mechanics and Mathematics of NAS of Ukraine, Naukowa Str. 3b, 79060 Lviv, Ukraine*

c *Department Life, Light & Matter, University of Rostock, Albert-Einstein-Strasse 25, 18059 Rostock, Germany*

d *Department of Chemistry, Swansea University – Singleton Park, Swansea SA2 8PP, United Kingdom*

* *Corresponding Authors: brindhu.seelan@uni-rostock.de, christian.klinke@uni–rostock.de*

**Abstract**

Quasi-2D metal halide perovskites have emerged as a promising material for photodetection due to excellent optoelectronic properties, simple synthesis, and robust stability. Albeit, developing high-performance photodetectors based on low-dimensional quasi-2D metal halide perovskite nanoparticles remains challenging due to quantum and dielectric confinement effects. Several approaches have been employed to improve efficiency, with plasmonic nanostructures being among the most effective ones. The resonant energy transfer and coupling between plasmons and excitons play a vital role in enhancing device performance. Here, we demonstrate enhanced photodetection of quasi-2D perovskite nanostripes resulting from the incorporation of octadecanethiol (ODT) functionalized Ag nanostructure arrays (ANA). Using colloidal lithography, ANA were fabricated. Reflectance spectroscopy and finite element method (FEM) simulations show that ANA supports localised surface plasmon resonance (LSPR) modes that spectrally coincide with the absorption and emission band of the perovskite. This spectral overlap enables interesting coupling interactions between the excitons and plasmons. The ODT-functionalized ANA photodetectors exhibit weak to intermediate coupling, resulting in a photocurrent enhancement factor of 838 %. They achieve photoresponsivities of up to 70.41 mA $W^{-1}$, detectivities of $1.48 \times 10^{11}$ Jones and external quantum efficiencies of 21.55 %, which are approximately 10 times higher than those of the reference photodetector. We present an approach to optimize the plasmon-exciton coupling and non-radiative energy transfer for developing high-performance plasmonic-perovskite hybrid photodetectors.

# Introduction

Photodetectors, an essential optoelectronic device that converts optical signals to electrical signals, have attracted enormous research interest due to their wide range of applications in various fields such as digital imaging, optical communication, surveillance and sensing. [1–4] The growing demand for ultra-compact, cost-effective, low-power, lightweight and flexible devices has driven the exploration beyond traditional active materials (crystalline semiconductors such as Si, Ge, GaAs). [4–7] Low-dimensional organic-inorganic hybrid 2D layered perovskites are highly promising materials for photodetection due to straightforward syntheses, outstanding optoelectronic properties and enhanced stability in ambient conditions. This enhanced stability is attributed to natural passivation by hydrophobic organic ligands, making them more suitable for commercial application compared to their bulk form. [5,8,9] The most widely studied 2D layered perovskites are of the Ruddlesden-Popper (RP) type with a crystal structure $A`_2 A_{n-1} B_n X_{3n+1}$, where $A`$ is a long chain cation, A is a smaller organic cation ($MA^+$), B is a divalent metal cation ($Pb^{2+}$), X is halide anion ($Br^-$) and $n$ refers to the number of BX layers between neighboring spacers ($n$ = 1 to ∞ (bulk)). [8–10] In this structure, the multiple quantum and dielectric confinements are formed due to $BX_6$ blocks separated by an alkyl chain. The quantum and dielectric confinements hinder photodetection due to the increased exciton binding energy, and anisotropic charge transport, restricting carrier movement within the quantum well plane. [11,12] Low-dimensional perovskites exhibit higher absorption coefficients compared to bulk. However, their limited volume as an active material in photodetection reduces overall light absorption, while the insulating hydrophobic passivation layer further limits conductivity. [8] Quasi-2D perovskite materials are considered intermediate materials exhibiting reduced quantum and dielectric confinement, resulting in improved carrier generation, mobility, diffusion, reduced trap states and stability compared to purely 2D perovskite. [13–15]

The photodetection efficiency of perovskite-based photodetectors can be improved by incorporating plasmonic metal nanostructures, utilizing the interaction of plasmons and excitons. [16–18] The metal nanostructures generate surface plasmons (SP), which are collective oscillations of the metal's electron density induced by the incident electromagnetic light at the metal-dielectric interface. [19] LSPR are confined modes that enable subdiffraction-limit light confinement, generating enhanced electric fields near the metal nanostructure. This enhanced electric field boosts light absorption by several orders of magnitude in neighbouring perovskite materials via near-field effects. [20] Additionally, the far-field enhancement through scattering of electromagnetic fields increases the photon interaction range, contributing to light absorption. [21,22] The decay of SP effectively generates hot carriers, which can be injected into adjacent perovskite material via a hot electron injection (HEI) process, enhancing the photocurrent. [16] In the near field, the plasmonic nanostructures and the perovskite can exchange energy via a non-radiative energy transfer process, such as plasmon-induced resonance energy transfer (PIRET). [23–26] Despite the enhancement of photodetection efficiency by plasmons, achieving optimum coupling between light and matter remains a challenge. To develop an efficient plasmon-enhanced perovskite photodetector, it is essential to study and optimize the optical coupling mechanisms between plasmons and excitons. These light-matter interactions can be tuned by controlling the size, shape and spatial arrangement of plasmonic nanostructures, as well as their proximity to the perovskite material. [20]

In this work, we report enhanced photodetection efficiency in quasi-2D $(C_{12}H_{27}N)_2(MA)_{n-1}(Pb)_n(Br)_{3n+1}$ perovskite nanostripes through the incorporation of ODT-functionalized ANA fabricated using a simple and cost-effective colloidal lithography technique. The quasi-2D perovskite nanostripes were prepared through the colloidal hot-injection method. The absorption and PL studies of perovskite nanostripes reveal that they consist of mixed-n phases and exhibit weaker quantum confinement dominated by high-n phases. ANA exhibits two

LSPR modes, and the resonance around 525 nm contributes to the photocurrent enhancement due to its spectral overlap with the perovskite's absorption and emission. The steady-state and time-resolved PL measurements were performed to determine the energy transfer mechanism and coupling strength between plasmons and excitons. The perovskite nanostripes show a strong coupling regime when in direct contact with ANA, while ODT-functionalized ANA show weak or intermediate coupling. The photocurrent significantly improves for perovskite nanostripes on ODT-functionalized ANA with a maximum photocurrent enhancement factor of 838 %. This study provides strategies to optimize light-matter interactions for the development of cost-effective, high-performance plasmonic perovskite hybrid photodetectors.

**Experimental section**

**Chemicals and reagents:** All chemicals were used as received. Polystyrene (PS) particles with a diameter of 0.5 μm (2 wt% dispersion in water, polydispersity 2.6 %) were obtained from Sigma Aldrich GmbH, Germany. 1-octadecanethiol (ODT, 98%, Sigma Aldrich). Lead (II) acetate tri-hydrate (Aldrich, 99.999%), nonanoic acid (Alfa Aesar, 97%), tri-octylphosphine (TOP; ABCR, 97%), 1-bromotetradecan (BTD; Aldrich, 97%), methylammonium bromide (MAB; Aldrich, 98%), diphenyl ether (DPE; Aldrich, 99%), toluene (VWR, 99,5%), dimethylformamide (DMF; Aldrich, 99,8%), dodecylamine (DDA; Merck, 98%), Isopropyl alcohol (IPA; Sigma Aldrich, 99.5%), Methyl isobutyl ketone (MIBK; Sigma Aldrich), acetone (HPLC grade, LiChrosolv, Merck, Germany).

**Synthesis of $(C_{12}H_{27}N)_2(MA)_{n-1}(Pb)_n(Br)_{3n+1}$ nanostripes:** The quasi-2D nanostripes are synthesized through the colloidal hot injection method as previously reported. [10] Briefly, A three-neck 50 mL flask was used with a condenser, septum, and thermocouple. 6 mL of diphenyl ether (37.7 mmol), 0.08 mL of a 500 mg dodecylamine (2.69 mmol) in 4 mL diphenyl ether precursor were heated to 80 °C in a nitrogen atmosphere. At 80 °C 0.2 mL of trinoctylphosphine (0.45 mmol) was added. Then vacuum was applied to dry the solution. After 1.5 h, the reaction apparatus was filled with nitrogen once again and the temperature was set to 130 °C. Then 0.8 mL of $PbBr_2$ nanosheets in toluene, synthesized as previously [10] were added and heated until everything was dissolved. The synthesis was started with the injection of 0.05 mL of a 300 mg methylammonium bromide (2.68 mmol) solution in 6 mL of dimethylformamide precursor. After 10 minutes, the heat source was removed and the solution was left to cool down below 60 °C. Afterward, it was centrifuged at 4000 rpm for 3 minutes. The particles were washed two times in toluene before the product was finally suspended in toluene again.

**Fabrication of the reference and plasmonic-perovskite hybrid photodetectors:** Figure 1 (a-i) illustrates the fabrication process of reference (perovskite/ODT) and plasmonic-perovskite hybrid (perovskite/ANA and perovskite/ODT/ANA) photodetectors. The $SiO_2$/Si (300 nm $\pm$ 5% thermal oxide, resistivity of .005-.020 Ω-cm, Addison Engineering) substrates of dimension 1 × 1 cm were washed by sonicating in acetone and IPA for 2 min in an ultrasonic bath, followed by drying with nitrogen gas. The interdigitated silver electrodes with an electrode spacing of 5 μm were fabricated on $SiO_2$/Si substrates using electron beam lithography. A 50 μL e-beam resist (ARP632) was spin-coated onto the substrate at 4000 rpm for 60 s and baked at 150 °C for 3 minutes. Electron beam exposure and development in a MIBK/IPA (1:3) solution for 60 s, and then proceeding with thermal evaporation of Ag and lift off process produced interdigitated silver electrodes. These electrodes are surface functionalized with ODT by immersing them in a 2 mM solution in ethanol for 2 h. [27] Based on the chemical functional group, ODT selectively binds only to the silver electrodes. The unbounded (physisorbed) ODT is removed by soaking for 30 minutes and rinsing thrice in ethanol and drying in nitrogen gas. The 20 μL solution of colloidal perovskite nanostripes in toluene is drop-casted on ODT functionalized interdigitated Ag electrodes to form a reference photodetector as shown in Figure 1 (g). To fabricate a plasmonic-perovskite hybrid photodetector (as illustrated in Figure 1 (a-f)), the electrode substrate was treated with oxygen plasma for 1 min to render surface hydrophilicity, which is necessary for the self-assembly of PS particles in a hexagonal close-packed (hcp) arrangement. The oxygen plasma treatment leads to the formation of perforated Ag electrodes due to etching. [28] The ANA was fabricated using the colloidal lithography reported earlier. [29] Briefly, a PS monolayer is formed using the evaporation-induced convective self-assembly technique. Followed by Ag deposition of thickness 35 nm with an adhesion promoter thin film of Ti (2 nm). Then, PS particles were removed by immersing in methylene chloride ($CH_2Cl_2$), resulting in ANA within the electrode gaps. The 20 µL solution of colloidal perovskite nanostripes in toluene is drop-casted on ANA-interdigitated Ag electrodes to form an ANA photodetector as shown in Figure 1 (h). The ODT/ANA photodetector, illustrated in Figure 1 (i), is fabricated by surface functionalizing the ANA structures with ODT, followed by drop-casting a 20 µL solution of colloidal perovskite nanostripe in toluene.

**Photodetector Characterization:** The I-V characteristics of the device under various illumination power densities and bias voltages were measured using a Keithley 4200 semiconductor characterization system. A laser with an excitation wavelength of λ = 405 nm was selected to simultaneously excite all the mixed n-phases in the perovskite. The photodetector's transient current response was measured under different bias voltages and power densities. A function generator (RIGOL, DG4062) was used to modulate the laser light into pulsed light for response time measurements. The current frequency response was measured with pulsed laser light and an oscilloscope (Tektronix, TDS2014B) at a bias voltage of 2 V (SRS70, power supply) and power density of 5.2 mW/cm$^2$.

**Characterization:** The TEM samples were prepared by diluting the nanostripe suspension with toluene, followed by drop casting 10 µL of the suspension on a TEM copper grid coated with a carbon film. Standard images were done on a Talos-L120C and EM-912 Omega with a thermal emitter operated at an acceleration voltage of 120 kV and 100 kV.

UV-visible (UV-Vis) absorption and reflection spectra were obtained with a Lambda 1050+ spectrophotometer from Perkin Elmer equipped with an integration sphere. The photoluminescence (PL) spectra measurements were obtained by a fluorescence spectrometer (Spectrofluorometer FS5, Edinburgh Instruments). For the time-resolved photoluminescence (TRPL) measurements, a picosecond laser with 375 nm excitation wavelength and 100 kHz repetition rate was used. The decay profiles are tail-fitted with a bi-exponential function $R(t) = A_1 \exp\left(-\frac{t}{\tau_1}\right) + A_2 \exp\left(-\frac{t}{\tau_2}\right)$ and the average PL lifetime is calculated using the formula $\tau_{average} = \frac{A_1\tau_1^2 + A_2\tau_2^2}{A_1\tau_1 + A_2\tau_2}$, where $A_1$ and $A_2$ are amplitude coefficients and $\tau_1$ and $\tau_2$ are the lifetimes of fast and slow decay components, respectively.

Photoluminescence quantum yield (PLQY) of the sample was measured using an absolute method by directly exciting the sample and the reference (toluene) solution in an SC-30 integrating sphere module fitted to a Spectrofluorometer FS5 from Edinburg Instruments. The calculation of absolute PL QY is based on the formula, $\eta = \frac{E_{sample} - E_{ref}}{S_{ref} - S_{sample}}$, where $\eta$ is absolute PL QY, $E_{sample}$ and $E_{ref}$ are the integrals at the emission region for the sample and the reference, respectively, and $S_{sample}$ and $S_{ref}$ are the integrals at the excitation scatter region for the sample and the reference, respectively. The selection and calculation of integrals from the emission and excitation scattering region and the calculation of absolute PLQY were performed using the FLUORACLE software from the Edinburg Instrument.

The confocal microscope platform Micro Time 200 (PicoQuant) was used for spatially resolved PL and fluorescence lifetime measurements using time-correlated single photon counting (TCSPC) method. A fibre-coupled picosecond laser with a wavelength of 442 nm and repetition rate of 40 MHz was used for excitation and focused (100X, Olympus air corrected) on a drop-cast sample. The emitted light was guided through a 465 nm long pass filter to the detection path, which was easily switchable by a flipable mirror between fluorescence lifetime (single photon detector, PMA Hybrid) and PL spectra (Andor Kymera SR193). Spatially resolved measurements were obtained by a galvo scanner unit (FLIMbee, PicoQuant). Regions of interest could be accessed by a piezo-motor driven x-y-z table (PI).

A scanning electron microscope (SEM, Zeiss EVO/MA10) was used for imaging of deposited colloidal nanostripes on interdigitated silver electrodes and ANA. We used the Inbeam secondary electron detection mode at an operating voltage of 10 kV and a working distance of 10 mm. SEM images were analyzed using ImageJ software to calculate the active area for each device.

Finite element method (FEM) simulations were performed on hcp ordered ANA. The COMSOL Multiphysics 6.2 (wave optics module) solves 3D Maxwell's equations to obtain the plasmonic response of ANA in the visible range. The dielectric function of the Ag film is taken from Johnson and Christy for the visible range. [30] From SEM images, the periodicity and size were estimated for constructing the unit cell of hcp ANA. The model is constructed for ANA on a Si/SiO$_2$ substrate. A 400 nm supersubstrate is modelled on top of ANA. A perfect match layer (PML, 200 nm) is added to the top and bottom. It absorbs the outgoing light and avoids multiple reflections, interacting with the interior region. The periodic boundary condition is applied on two parallel sides (xz and yz planes) of the model. A linearly polarized wave (in x-direction) is set to strike normally on the ANA from the port. The field distribution was simulated as a normalized electric field (V/m). We have used a user-defined mesh, which consists of a fine mesh for the small sub-domain and a normal mesh for the rest of the areas.

## Result and Discussion

The fabrication of a plasmonic-perovskite hybrid and reference photodetectors is illustrated in Figure 1 (a-i), with corresponding SEM images at various fabrication stages in Figure 1 (j-m). The fabrication steps are described in the Experimental section. Figure 1 (j) shows the perforated Ag electrodes obtained post-oxygen plasma treatment of interdigitated Ag electrodes. The oxygen plasma treatment renders the surface hydrophilic, which is essential for the self-assembly of the PS particles into an hcp arrangement as shown in Figure 1 (k). The oxygen plasma etches the Ag layer, resulting in the formation of the perforated electrodes. The hcp PS monolayer serves as a template during Ag metal deposition, and upon subsequent removal of the PS, ANA structures form within the electrode gaps as shown in Figure 1 (l). The size of the Ag triangles is around $130 \pm 13$ nm, as shown in Figure S1 (a, b). Figure 1 (m) shows the fabricated plasmonic-perovskite hybrid surface after the deposition of quasi-2D colloidal perovskite nanostripes onto the ODT-functionalized ANA. The ODT self-assembled monolayer (SAM) serves as a separator, improving chemical stability. [31] Figure S2 shows the SEM images of the plasmonic-perovskite hybrid and reference photodetector.

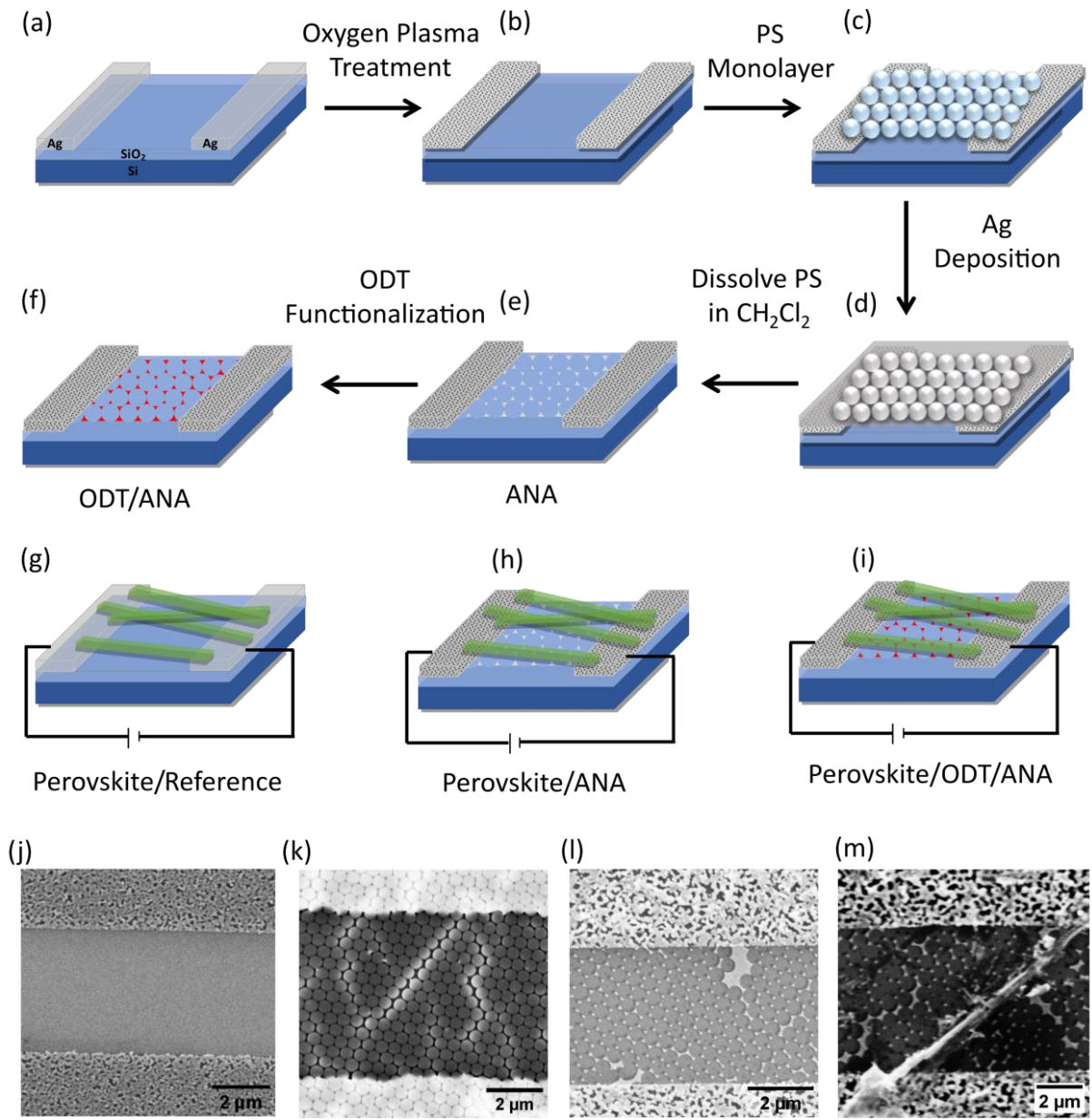

**Figure 1.** (a-i) Schematic illustration of the fabrication process for a plasmonic-perovskite hybrid and reference photodetector. (j-m) SEM images corresponding to various stages of the fabrication process. (j) Perforated Ag electrodes formed on SiO$_2$/Si substrate as a result of oxygen plasma treatment. (k) SAM of PS particles formed on perforated Ag electrodes. (l) ANA fabricated within the electrode gaps. (m) Quasi-2D perovskite nanostripes deposited on ODT-functionalized ANA.

The bright field TEM images of quasi-2D perovskite nanostripes synthesized through the colloidal hot injection method with slight modifications of the precursor concentration, reaction temperature, and duration are shown in Figure 2 (a). [10] The nanostripes exhibit an elongated rectangular shape with lengths ranging from 1 μm to 7 μm and widths between 12 - 250 nm. Figure 2 (b) shows the X-ray powder diffraction (XRD) characterization for nanostripes. The strong reflexes (100) and (200) are observed and assigned to the crystallographic planes of bulk methylammonium lead bromide (MAPbBr$_3$).[32] The inset shows small repeating reflexes marked as (002) to (008), confirming the RP quasi-2D structure. Additionally, low-intensity broad peaks (such as 7.4) support the presence of mixed n-phases.

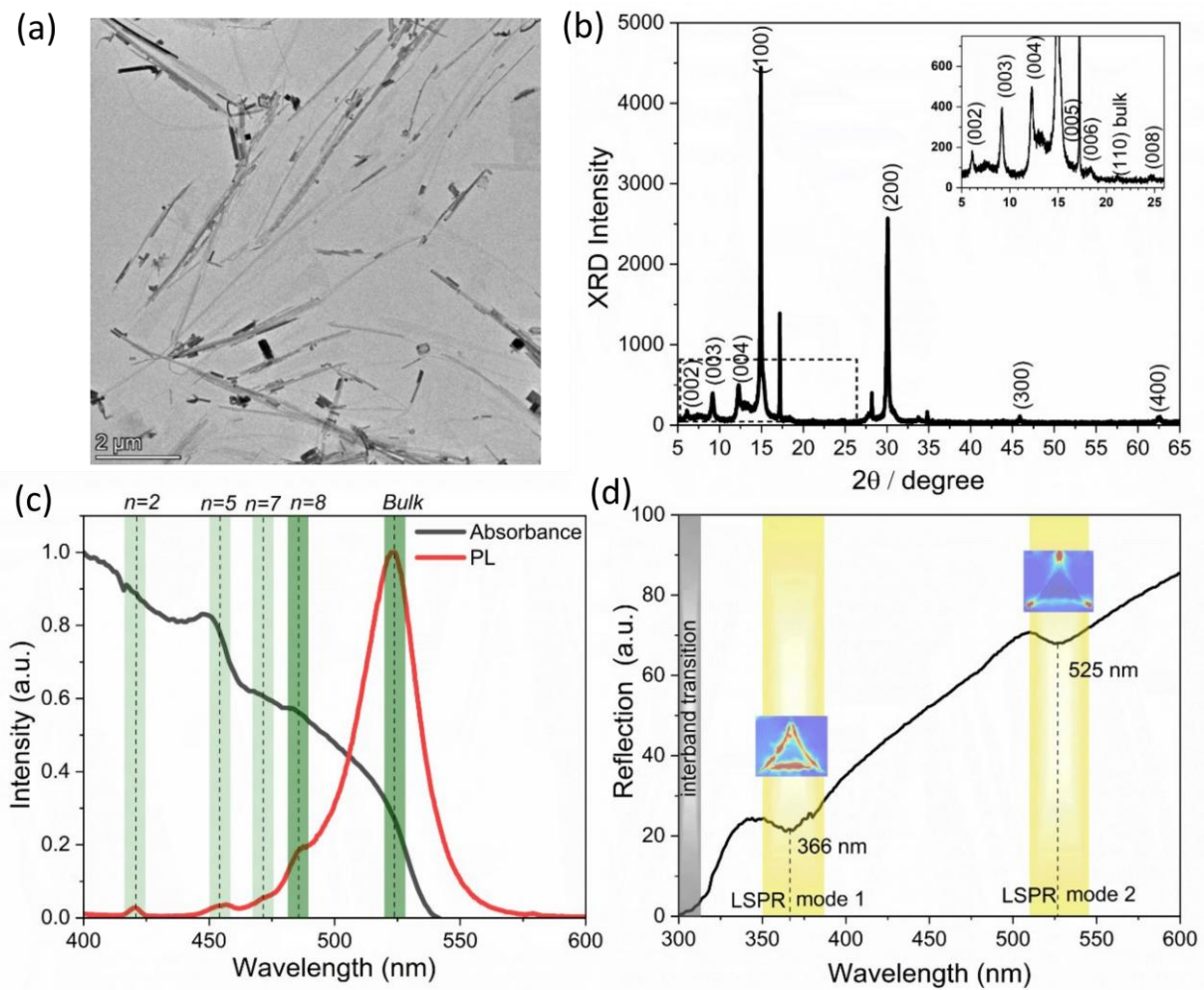

**Figure 2.** (a) TEM images of quasi-2D perovskite nanostripes (b) XRD analysis of quasi-2D perovskite nanostripes. The inset shows the zoomed-in pattern indicating the presence of mixed-n phases. (c) Absorbance and PL spectra of quasi-2D perovskite nanostripes, with vertical lines indicating different n-phases. The high-n phases are shown in dark green and the low-n phases in light green. (d) Reflection spectrum of ANA. The spectral features, marked by dashed lines, represent LSPR modes. The inset shows the corresponding FEM simulation of the electric field distribution for the Ag triangle.

The different n-phases, in combination with the bulk phase, are recognizable in the steady-state UV-visible absorbance and PL spectroscopy, as shown in Figure 2 (c). The multiple absorbance peaks as listed in Table S1 are attributed to different phases $n$ = 2, 5, 7, 8 and ∞ respectively. [12,33] A prominent spectral feature for $n$ = 5, along with bulk phases, suggests that the nanostripes are mostly composed of high-n phases and exhibit weaker confinement. Nanostripes exhibit a significant excitonic effect in UV-Vis and its contribution can be extracted using Elliott fitting as reported earlier. [34] The extracted excitonic peak position (523 nm [2.37 eV]) along with band gap and binding energy is shown in Figure S3. The PL spectra exhibit strong emission at around 524 nm and a weak emission at lower wavelengths originating from the confinement effect due to low-n phases. The TRPL profile (Figure S4) with its fitting parameters is listed in Table S2. The nanostripes in solution show an average lifetime of 9.6 ns with a PLQY of 1.36 %. Figure 2 (d) shows the reflection spectrum of the ANA at normal incidence. The reflectance spectrum exhibits two dips around 366 nm and 525 nm, attributed to LSPR modes. The incident light is resonantly absorbed or scattered by the ANA, resulting in a dip in the reflection spectrum. The dip around 317 nm is attributed to the interband electronic transition from the d-band to the conduction band. [30] Figure S5 shows the simulated reflectance spectrum with the electric field distribution (V/m) for different LSPR modes. The experimental LSPR modes are blueshift compared to the simulation. This discrepancy may be attributed to minor imperfections intrinsic to colloidal lithography, such as defects, grain boundaries, and size variation, as shown in Figure S1 (b). [35] At 366 nm (LSPR mode 1), a strong localized electric field around the rim of the Ag triangle at the air/Ag interface is observed. At 525 nm (LSPR mode 2), the electric field is concentrated and localized (hot spots) at the tips of the Ag triangle at the air/Ag interface. The spectral overlap between the LSPR mode 2 (around 525 nm) and perovskite's absorbance and PL emission enables interesting interactions between the plasmon and exciton, depending on the coupling regime (strong, intermediate, or weak). The strong coupling results in Rabi splitting and the formation of plexciton quasiparticles. [36] The intermediate and weak coupling results in plasmon-induced resonance energy transfer (PIRET) and the Purcell effect, enabling the possibility to control the effective radiative lifetime of charge carriers. [37] The effective energy transfer between plasmons and excitons is crucial for enhancing photodetection performance.

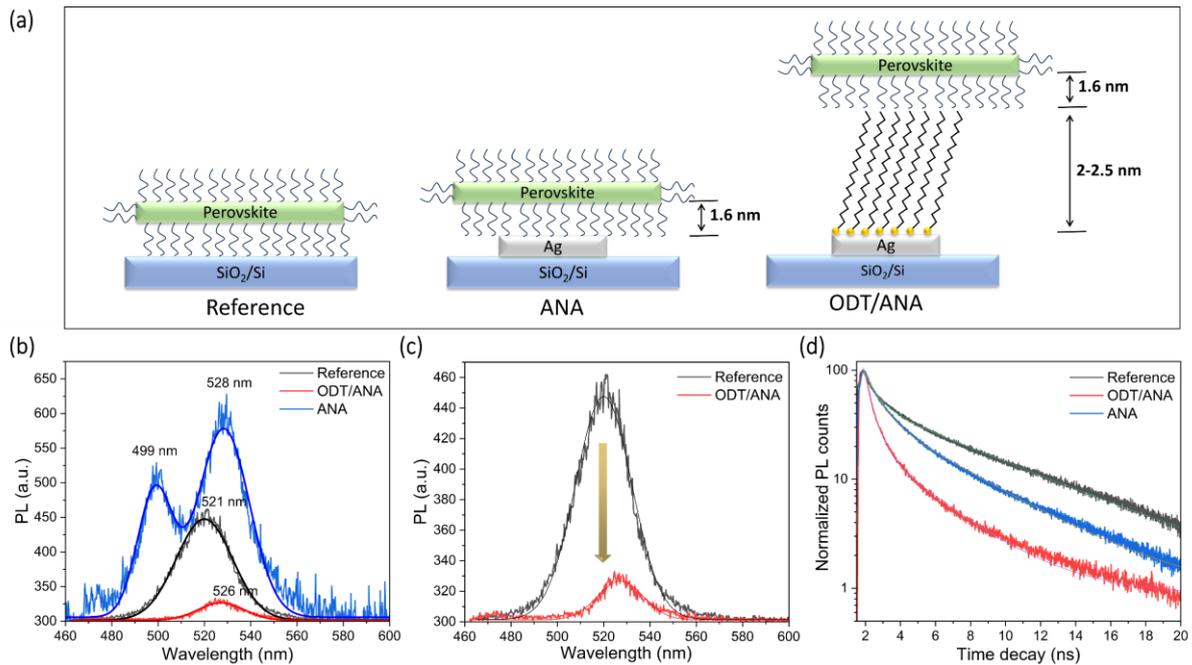

**Figure 3.** (a) Schematic illustration of different substrates used for PL measurements. (b, c) PL spectra of perovskite nanostripes on various substrates. (b) PL spectra of perovskite on ANA substrate exhibit Rabi splitting. (c) The reduction in the intensity of PL spectra is observed on the ODT/ANA substrate compared to the reference substrate. (d) TRPL profiles excited with a pulse laser of 442 nm for perovskite nanostripes on different substrates.

To understand plasmon exciton coupling and the dependence of energy transfer on the proximity to ANA, steady-state and time-resolved PL measurements were carried out on three distinct substrates as schematically shown in Figure 3 (a). The perovskite nanostripes are passivated using the spacer-ligand dodecylamine (DDA), which has an approximate length of 1.6 nm. The ODT forms SAM on Ag with a thickness ranging from 2 to 2.5 nm. [38] Figure 3 (b) shows the PL spectra of perovskite on different substrates. The perovskite deposited on the reference substrate exhibits a PL emission peak at around 521 nm, showing a slight blueshift of 3 nm compared to its emission in the solution. This shift can be attributed to different measurement conditions and changes in the dielectric environment. In solution, PL is measured from an ensemble of nanostripes (broad size distribution) in toluene, while on (SiO$_2$/Si) substrate, it is measured from a single nanostripe. The PL spectra on the ANA substrate significantly differ and exhibit Rabi splitting with two distinct peaks at 499 nm and 528 nm corresponding to the high and low energy branches due to the overlap of exciton and LSPR energy. From the coupled harmonic oscillator model the high ($E_+$) and lower energy branch ($E_-$) are described as $E_\pm = \frac{1}{2}(E_P + E_e) \pm \frac{1}{2}\sqrt{4g^2 + \delta^2}$, where $E_P$ and $E_e$ are plasmon and exciton energies, $\delta = E_P - E_e$ is the energy detuning, $g$ is the coupling strength and $2g$ gives the Rabi splitting energy at zero detuning. [37,39] The ANA substrate exhibits a Rabi splitting energy of 139 meV. Additionally, the strong coupling should satisfy the coupling strength and damping rate criterion $2g > \frac{\gamma+\kappa}{2}$, where γ (80 meV) and κ (71 meV) are the damping rates of plasmon and exciton, respectively, indicating strong coupling. When the spacing between perovskite and ANA is increased by ODT functionalization, a slightly redshifted PL peak without Rabi splitting is observed, suggesting an intermediate or weak coupling regime. In this regime, the PIRET mechanism should dominate over strong coupling. On the ODT-functionalized ANA substrate, the PL peak intensity decreases by 82 % compared to the reference substrate, as shown in Figure 3 (c). The suppression of the PL is attributed to possible energy transfer from the perovskite to ANA. [40,41] Figure 3 (d) shows the TRPL recorded for perovskites on different substrates under excitation by 442 nm (above the bandgap energy of perovskite) and the fitting parameters are listed in Table S2. The average lifetime decreases in both ANA and ODT/ANA substrates compared to the reference sample in the order SiO$_2$/Si (3.61 ns) > ANA (2.18 ns) > ODT/ANA (0.64 ns). The LSPR increases the radiative decay rate, leading to shorter lifetimes after incorporating ANA. As the proximity between the perovskites and the plasmonic nanostructures decreases, the lifetime typically shortens. [42,43] The suppressed PL intensity with reduced lifetime observed in ODT/ANA suggests possible energy transfer between the perovskite and ANA. In particular, the shortened $\tau_1$ in ODT/ANA further supports the presence of energy transfer. [44,45]

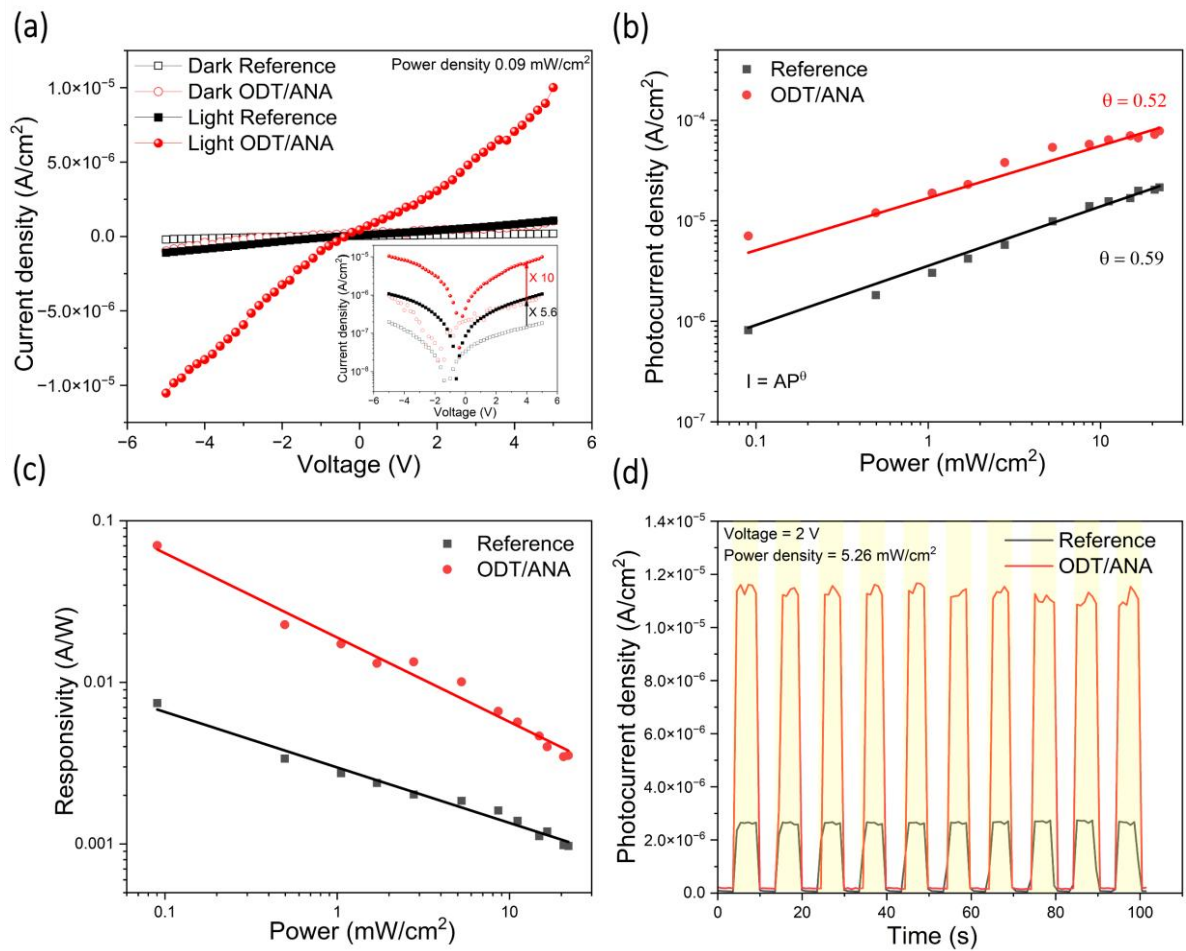

**Figure 4.** (a) Current density-voltage (I-V) measurements for reference and ODT/ANA photodetector using an illuminating laser (λ = 405 nm) with a power density 0.09 mW/cm² in the range -5 to +5 V. The inset shows a logarithmic I-V response, where the ODT/ANA photodetector exhibits a photocurrent 10 times its dark current, while the reference photodetector shows 5.6 times. (b) Photocurrent measured under varying illumination power at a bias voltage of 4 V. The photocurrent is expressed by the power exponent law $I = AP^\theta$, where $A$ is a constant and $\theta$ defines the response of the photocurrent density to the light power. (c) Responsivity evaluated with different illumination power densities at a bias of 4 V. (d) Transient photocurrent density measured at constant bias voltage 2 V and illumination power 5.2 mW/cm⁻².

The fundamental mechanism of photodetection involves the generation and extraction of photoinduced carriers. The photodetection performance of the reference and plasmonic-perovskite hybrid devices (ANA and ODT/ANA) was measured to investigate the influence on these processes due to the incorporation of ANA. The plasmonic-perovskite hybrid devices (ANA) were fabricated by depositing the perovskite nanostripes on the ANA and ODT-functionalized ANA patterned substrate. The current-voltage (I-V) measurements for the ANA substrate show a decrease in photocurrent with light illumination, as shown in Figure S6. This anomalous behaviour could be due to interfacial chemical interaction between Ag and perovskite, where Br ion migration leads to the formation of AgBr, degrading ANA and contacts. [46] Additionally, LSPR can induce thermal effects that accelerate perovskite decomposition and suppress charge extraction. [47,48] This issue is addressed by functionalizing the ANA substrate with an ODT monolayer, which serves as a protective interfacial layer that inhibits chemical interaction.

Figure 4 (a) compares current-voltage (I-V) measurements for ODT/ANA and reference photodetectors under dark and light conditions (power density 0.09 mW/cm$^2$). The photocurrent density of $7.06 \times 10^{-6}$ A/cm$^2$ for the ODT/ANA substrate and $8.18 \times 10^{-7}$ A/cm$^2$ for the reference substrate are observed at a bias voltage of 4 V. The ODT/ANA substrate exhibits a photocurrent density that is 10 times higher than its dark current whereas the reference substrate show a photocurrent density, 5.6 times its dark current. The photocurrent enhancement factor ($\eta$) is given as $\eta = \frac{I_{ODT/ANA} - I_{reference}}{I_{reference}}$ (%) , where $I_{reference}$ and $I_{ODT/ANA}$ are the photocurrents of the reference and ODT/ANA substrates. [49] The photocurrent enhancement factor reaches a maximum value of 838 % at 5 V. Further, the dependence of the photocurrent on illumination power densities at 4 V is compared as shown in Figure 4 (b). The photocurrent increases with the illumination power densities for both the substrates, which can be described by the power exponent law $I = AP^\theta$, where $A$ is the constant and $\theta$ defines response of the photocurrent density to the light power. It gives $\theta$ = 0.52 and 0.59 for the reference and the ODT/ANA substrate, respectively as shown in Figure 4 (b). The ideal photodetector exhibits $\theta$ close to 1. As the measured value is 0.5< $\theta$ <1, it indicates the existence of defects and traps leading to complex electron-hole generation and recombination processes as reported earlier. [50] The photocurrent enhancement factor decreases with illumination power densities and exhibits maximum value of 762 % at the power density of 0.09 mW/cm² and minimum value of 235 % at the power density of 16.5 mW/cm² at 4 V. The enhanced photocurrent in ODT/ANA substrate should be attributed to the intermediate or weak coupling between the excitons and LSPR, which results in an PIRET process from ANA towards perovskite resulting in a massive generation of carriers (electron-hole pairs). In the range 0.5 to 5.3 mW/cm², the photocurrent increases linearly with the laser power density for both substrates, as shown in Figure S7. The linear dynamic range ($LDR$) given by $LDR = 20 \log \left(\frac{I_{max}}{I_{min}}\right)$, where $I_{max}$ and $I_{min}$ are the maximum and minimum photocurrent in the linear regime, respectively. The ODT/ANA and reference substrates show similar $LDR$ with 13 dB and 14 dB, respectively.

To further compare the photodetection performance of ODT/ANA and the reference substrate, the photodetector parameters, responsivity ($R$), external detectivity ($D$) and quantum efficiency ($EQE$) were evaluated using the expressions, $R = \frac{I_{photo} - I_{dark}}{PA}$, $D = \frac{R}{\sqrt{2e(I_{dark}/A)}}$, $EQE = \frac{Rhc}{e\lambda}$ where $I_{photo}$ and $I_{dark}$ are photocurrent and dark current, respectively. $P$ is the illumination light power density, $A$ is the active area of the photodetector, $h$ is Planck constant, $c$ is the speed of light, $e$ is the elementary charge and $\lambda$ is the wavelength of the laser. [51] The responsivity evaluated with different illumination power densities at a bias of 4 V for ODT/ANA and the reference photodetector is shown in Figure 4 (c). At the lowest power density, the highest $R$ of 7.46 mA W$^{-1}$ and 70.41 mA W$^{-1}$ are obtained for reference and ODT/ANA, respectively, and it declines with further increase in power. The ODT/ANA substrate exhibits higher *R* compared to the reference, with enhancement ranging from 9.4 to 3.3 times at different powers. Similar trends are observed in *D* and *EQE* evaluated at different power densities as shown in Figure S8 (a, b). At higher illumination powers, the reduced enhancement in parameters is attributed to the thermal effect generated from LSPR, which can induce increased non-radiative recombination and local degradation in the perovskite. [47,48] The reference and ODT/ANA substrate exhibit an *EQE* of 2.28 % and 21.55 %, respectively, along with a corresponding *D* of $3.46 \times 10^{10}$ Jones and $1.48 \times 10^{11}$ Jones. The enhancement in *D* in ODT/ANA ranges from 4.29 to 1.52 times that of the reference with power. The dependence of the photodetection parameters with voltage at a constant power density of 0.09 mW cm$^{-2}$ for ODT/ANA and reference is shown in Figure S8 (c, d). The ODT/ANA exhibits higher values than the references. The transient photocurrent density measurements at different bias voltages (at constant power of 5.2 mW/cm$^{-2}$) and different powers (at constant voltage of 2 V) were performed by switching the laser (405 nm) on and off with a period of 10 s, as shown in Figure S9. A higher photocurrent is observed with increasing voltage and laser power, attributed to enhanced carrier generation. Figure 4 (d), compares the transient photocurrent density of ODT/ANA with reference photodetector, showing ODT/ANA exhibits 4.35 times higher on/off ratios at 2 V bias.

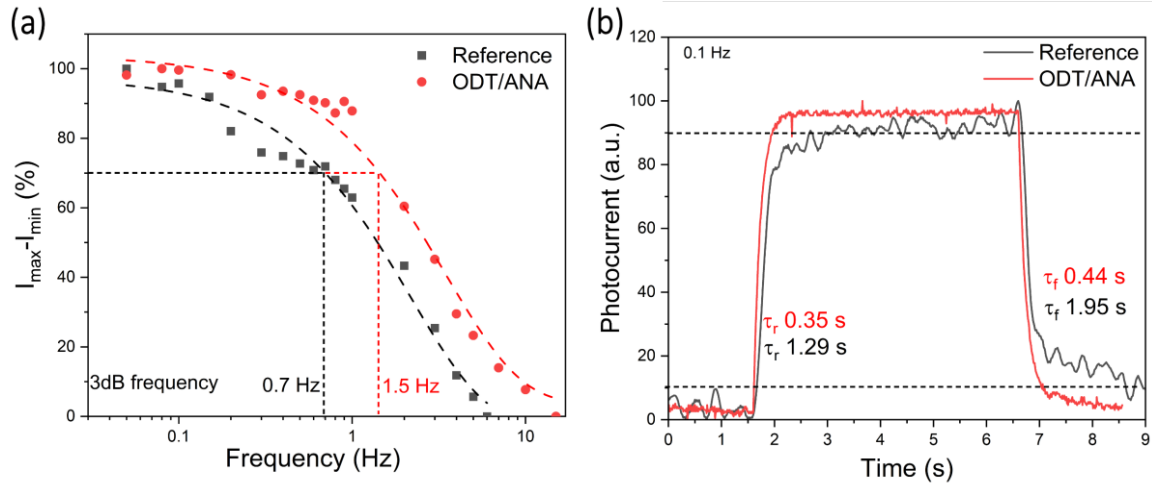

**Figure 5.** (a, b) Photoresponse characteristics at 2 V bias and pulse light illumination power 5.2 mW cm$^{-2}$ (a) temporal response and 3dB bandwidth at different frequencies. (b) Estimating rise time ($\tau_r$) and fall time ($\tau_f$) with rising and falling edges for reference and ODT/ANA at 0.1 Hz.

The frequency response of the normalized relative photocurrent ($I_{max}$-$I_{min}$) in the range 0.05 to 15 Hz was measured as shown in Figure 5 (a). The frequency at which the photodetector's response falls to 70.7% of its maximum value gives the 3dB frequency. The perovskite nanostripes exhibit a higher 3dB frequency of 1.5 Hz on the ODT/ANA substrate. Figure 5 (b) compares the response time. The rise time ($\tau_r$) and fall time ($\tau_f$) obtained for the reference are 1.29 s and 1.95 s, respectively. The perovskite on ODT/ANA shows a faster response time $\tau_r$=0.35 s and $\tau_f$=0.44 s than the reference. Figure S10 (a, b) shows the response time evaluated at 3dB frequency, showing similar behaviour. The faster response observed in ODT/ANA is attributed to LSPR coupling to perovskite, which generates and transports carriers effectively due to resonant energy transfer.

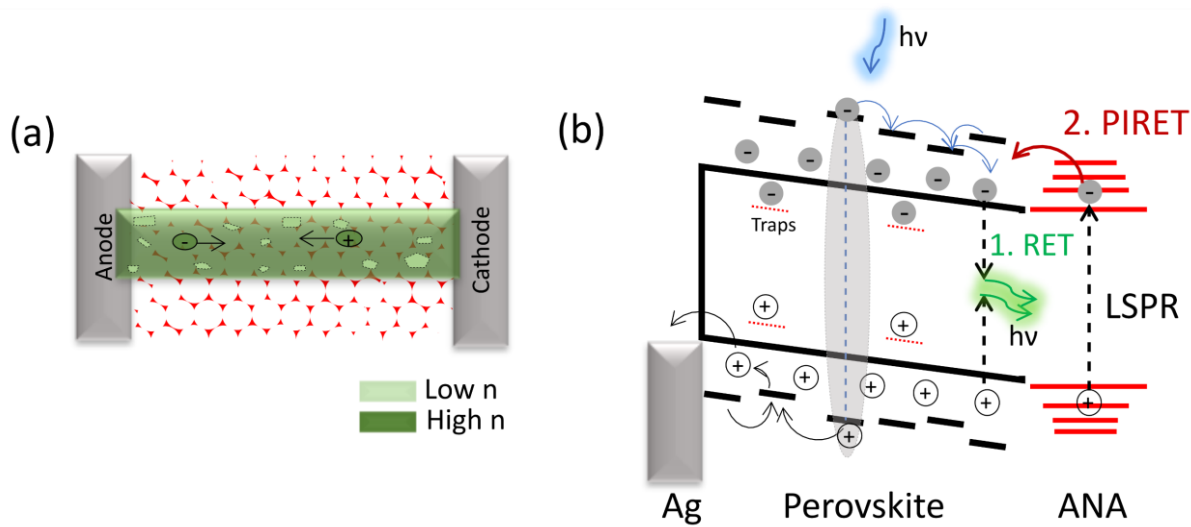

**Figure 6.** Schematic illustration of charge carrier transport and separation in a nanostripe deposited on ODT/ANA substrate. The individual nanostripes have mixed-n phases, mostly consisting of high-n phases (shown in dark green) with small fractions of low-n phases (shown in light green). (b) Mechanism of energy and charge transport in perovskite/ODT/ANA photodetector. The photogenerated excitons dissociate to free charge carriers and are separated by the Schottky barrier. 1. A fraction of photogenerated carriers undergo radiative recombination, exciting plasmons in ANA via RET. 2. Through dipole-dipole interaction via PIRET, plasmons excite charge carriers back into the perovskite.

The enhanced photodetection performance by the nanostripes on ODT/ANA substrate is illustrated by the charge transport mechanism schematics shown in Figure 6. The 405 nm laser light excites all the n-phases simultaneously in the perovskite, generating charge carriers (excitons and free charge carriers). Part of photogenerated excitons dissociates to free charge carriers through relaxation to the band-edge through the funneling process [33] and can recombine and emit at 524 nm. At the same time, other photogenerated carriers are separated by the electric field of the Schottky barrier. The spectral overlap between the perovskite emission and the ANA's LSPR at around 525 nm enables photons emitted by the perovskite to be absorbed by ANA through resonance energy transfer (RET). Additionally, the LSPR in ANA also overlaps with the perovskite's absorption peak (524 nm), facilitating absorption energy transfer via a non-radiative PIRET mechanism, which generates a large number of carriers back in the perovskite, contributing to the photocurrent. The SAMs of ODT monolayers typically form with some degree of disorder, including defects and pinholes, which can facilitate hot electron injection (HEI) due to the reduced insulating barrier between perovskite and ANA. [52] We speculate that HEI pathway is however unlikely to contribute significantly to the photocurrent due to low injection probability across the interface, with its impact being minor compared to the dominant PIRET pathway. In addition, the LSPR can enhance the local electromagnetic field in the near field regions of the plasmonic nanostructure, increasing the local photon density. [20] However, the suppressed PL intensity observed in Figure 3 (c) suggests that this effect does not notably contribute to the photocurrent enhancement. [53,54] The enhanced photodetection performance observed in nanostripes on ODT/ANA is primarily attributed to RET and PIRET mechanisms, with negligible contributions from HEI and increased local photon density.

## Conclusion

In summary, we demonstrated that incorporating ODT-functionalized ANA enhances the photodetection performance of perovskite nanostripes. The absorption and PL studies show that quasi-2D perovskite nanostripes are composed of mixed-n phases with a significant excitonic component. The ANA exhibits LSPR modes that generate localized enhanced electric fields at the tip of the Ag triangles. The spectral overlap between ANA and perovskite absorption and emission enables efficient coupling and energy transfer assisted by plasmons. The coupling between the plasmon and exciton is analysed using steady-state and time-resolved PL measurements. The nanostripes placed directly on ANA exhibit strong coupling, evidenced by Rabi splitting of 139 meV, whereas ODT-functionalized ANA show weak or intermediate coupling with a redshifted single PL emission peak. The ODT-functionalized ANA photodetector demonstrates enhanced photoresponsivity, detectivity and $EQE$ with values of 70.41 mA W$^{-1}$, $1.48 \times 10^{11}$ Jones and 21.55 % at 4 V bias, approximately 10 times higher than those of the reference photodetector. It exhibits the photocurrent enhancement factor of 838 % with faster photoresponse ($\tau_r$ = 350 ms and $\tau_f$ = 440 ms) and higher on/off ratio (4.35 times at 2 V bias). The enhanced photodetection is attributed to weak or intermediate coupling, which facilitates non-radiative energy transfer such as RET and PIRET. Our investigation contributes to the fundamental understanding of plasmon-exciton coupling in quasi-2D perovskite, which is necessary for developing cost-effective, high-performance plasmonic-perovskite hybrid photodetectors.

**Supplementary Information** (PDF)

**Author Contributions**

The manuscript was written through contributions of all authors. All authors have given approval to the final version of the manuscript.

**Conflicts of interest**

There are no conflicts to declare.

**Acknowledgments**

B.M.S. acknowledges Alexander von Humboldt-Stiftung/Foundation for the postdoctoral research fellowship. Deutsche Forschungsgemeinschaft (DFG, German Research Foundation) is acknowledged for funding of SFB 1477 "Light-Matter Interactions at Interfaces", project number 441234705, W03 and W05. C. K. also acknowledges the European Regional Development Fund of the European Union for funding the PL spectrometer (GHS-20-0035/P000376218) and X-ray diffractometer (GHS-20-0036/P000379642) and the DFG for funding an electron microscope Jeol NeoARM TEM (INST 264/161-1 FUGG) and an electron microscope Thermo Fisher Talos L120C (INST 264/188-1 FUGG).

**Abbreviations**

ANA, Ag nanostructure array; RP, Ruddlesden-Popper; TOP, tri-octylphosphine; BTD bromotetradecan; MAB, methylammonium bromide; DPE, diphenyl ether; DDA, dodecylamine; ODT, octadecanethiol; PL, photoluminescence; TRPL, time-resolved photoluminescence; PLQY, Photoluminescence quantum yield; SAM, Self-assembled monolayer; SEM, scanning electron microscope; PIRET, plasmon-induced resonance energy transfer; HEI, hot electron injection; LDR, linear dynamic range; R, responsivity; EQE, external quantum efficiency; D, detectivity.

# References

[1] F. P. García de Arquer, A. Armin, P. Meredith, E. H. Sargent, *Nat Rev Mater* **2017**, *2*, 16100.

[2] D. Wu, Y. Zhang, C. Liu, Z. Sun, Z. Wang, Z. Lin, M. Qiu, D. Fu, K. Wang, *Advanced Devices & Instrumentation* **2023**, *4*.

[3] Y. Wang, L. Song, Y. Chen, W. Huang, *ACS Photonics* **2020**, *7*, 10.

[4] T. Wang, D. Zheng, J. Zhang, J. Qiao, C. Min, X. Yuan, M. Somekh, F. Feng, *Adv Funct Mater* **2022**, *32*.

[5] H. Wang, S. Li, X. Liu, Z. Shi, X. Fang, J. He, *Advanced Materials* **2021**, *33*.

[6] C. Li, J. Li, Z. Li, H. Zhang, Y. Dang, F. Kong, *Nanomaterials* **2021**, *11*, 1038.

[7] Y. Cheng, X. Guo, Y. Shi, L. Pan, *Nanotechnology* **2024**, *35*, 242001.

[8] Y. Yue, N. Chai, M. Li, Z. Zeng, S. Li, X. Chen, J. Zhou, H. Wang, X. Wang, *Advanced Materials* **2024**, *36*.

[9] I. Metcalf, S. Sidhik, H. Zhang, A. Agrawal, J. Persaud, J. Hou, J. Even, A. D. Mohite, *Chem Rev* **2023**, *123*, 9565.

[10] E. Klein, A. Black, Ö. Tokmak, C. Strelow, R. Lesyuk, C. Klinke, *ACS Nano* **2019**, *13*, 6955.

[11] H. Gu, S. Chen, Q. Zheng, *Adv Opt Mater* **2021**, *9*.

[12] B. Malani S, E. Klein, R. Lesyuk, C. Klinke, *Adv Funct Mater* **2025**, *35*.

[13] Z. Wang, Q. Wei, X. Liu, L. Liu, X. Tang, J. Guo, S. Ren, G. Xing, D. Zhao, Y. Zheng, *Adv Funct Mater* **2021**, *31*.

[14] A. Yadav, S. Ahmad, *ACS Appl Mater Interfaces* **2024**, *16*, 43134.

[15] E. Klein, C. Rehhagen, R. Lesyuk, C. Klinke, *J Mater Chem C Mater* **2023**, *11*, 9495.

[16] J. Huang, L. Luo, *Adv Opt Mater* **2018**, *6*.

[17] F. Wang, X. Zou, M. Xu, H. Wang, H. Wang, H. Guo, J. Guo, P. Wang, M. Peng, Z. Wang, Y. Wang, J. Miao, F. Chen, J. Wang, X. Chen, A. Pan, C. Shan, L. Liao, W. Hu, *Advanced Science* **2021**, *8*.

[18] Y. Xi, X. Wang, T. Ji, G. Li, L. Shi, Y. Liu, W. Wang, J. Ma, S. (Frank) Liu, Y. Hao, L. Xiao, Y. Cui, *Adv Opt Mater* **2023**, *11*.

[19] W. L. Barnes, A. Dereux, T. W. Ebbesen, *Nature* **2003**, *424*, 824.

[20] D. Zheng, Y. Prado, T. Pauporté, L. Coolen, *Adv Opt Mater* **2025**, *13*.

[21] R. S. Nithyananda Kumar, A. Martulli, S. Lizin, W. Deferme, *Prog Mater Sci* **2025**, *153*, 101479.

[22] Z. Yao, Y. Xiong, H. Kang, X. Xu, J. Guo, W. Li, X. Xu, *ACS Omega* **2024**, *9*, 2606.

[23] J. Li, S. K. Cushing, F. Meng, T. R. Senty, A. D. Bristow, N. Wu, *Nat Photonics* **2015**, *9*, 601.

[24] A. Souzou, M. Athanasiou, A. Manoli, M. Constantinou, M. I. Bodnarchuk, M. V. Kovalenko, C. Andreou, G. Itskos, *ACS Photonics* **2025**, *12*, 2344.

[25] X. You, S. Ramakrishna, T. Seideman, *J Chem Phys* **2018**, *149*.

[26] S. K. Cushing, J. Li, J. Bright, B. T. Yost, P. Zheng, A. D. Bristow, N. Wu, *The Journal of Physical Chemistry C* **2015**, *119*, 16239.

[27]    H. Ron, S. Matlis, I. Rubinstein, *Langmuir* **1998**, *14*, 1116.

[28]    C. Yuan, D. Zhang, Y. Gan, *ACS Omega* **2024**, *9*, 28912.

[29]    B. Malani S, P. Viswanath, *Journal of the Optical Society of America B* **2018**, *35*, 2501.

[30]    P. B. Johnson, R. W. Christy, *Phys Rev B* **1972**, *6*, 4370.

[31]    O. Salihoglu, A. Muti, A. Aydinli, (Eds.: Andresen, B. F.; Fulop, G. F.; Hanson, C. M.; Norton, P. R.; Robert, P.), **2013**, p. 87040T.

[32]    G. A. Elbaz, D. B. Straus, O. E. Semonin, T. D. Hull, D. W. Paley, P. Kim, J. S. Owen, C. R. Kagan, X. Roy, *Nano Lett* **2017**, *17*, 1727.

[33]    A. Niebur, E. Klein, R. Lesyuk, C. Klinke, J. Lauth, *Adv Opt Mater* **2025**, *13*.

[34]    R. M. Piehler, E. Klein, F. M. Gómez-Campos, O. Kühn, R. Lesyuk, C. Klinke, *Adv Funct Mater* **2025**.

[35]    C. L. Haynes, R. P. Van Duyne, *J Phys Chem B* **2001**, *105*, 5599.

[36]    A. E. Schlather, N. Large, A. S. Urban, P. Nordlander, N. J. Halas, *Nano Lett* **2013**, *13*, 3281.

[37]    R. Su, A. Fieramosca, Q. Zhang, H. S. Nguyen, E. Deleporte, Z. Chen, D. Sanvitto, T. C. H. Liew, Q. Xiong, *Nat Mater* **2021**, *20*, 1315.

[38]    M. Jalal Uddin, M. Khalid Hossain, M. I. Hossain, W. Qarony, S. Tayyaba, M. N. H. Mia, M. F. Pervez, S. Hossen, *Results Phys* **2017**, *7*, 2289.

[39]    J. Wang, R. Su, J. Xing, D. Bao, C. Diederichs, S. Liu, T. C. H. Liew, Z. Chen, Q. Xiong, *ACS Nano* **2018**, *12*, 8382.

[40]    J. A. La, J. Kang, J. Y. Byun, I. S. Kim, G. Kang, H. Ko, *Appl Surf Sci* **2021**, *538*, 148007.

[41]    H. Wang, J. W. Lim, L. N. Quan, K. Chung, Y. J. Jang, Y. Ma, D. H. Kim, *Adv Opt Mater* **2018**, *6*.

[42]    Q. Shang, S. Zhang, Z. Liu, J. Chen, P. Yang, C. Li, W. Li, Y. Zhang, Q. Xiong, X. Liu, Q. Zhang, *Nano Lett* **2018**, *18*, 3335.

[43]    Q. Shang, S. Zhang, Z. Liu, J. Chen, P. Yang, C. Li, W. Li, Y. Zhang, Q. Xiong, X. Liu, Q. Zhang, *Nano Lett* **2018**, *18*, 3335.

[44]    M. Niu, C. Shan, C. Xue, X. Xu, A. Zhang, Y. Xiao, J. Wei, D. Zou, G. J. Chen, A. K. K. Kyaw, P. P. Shum, *ACS Appl Mater Interfaces* **2025**, *17*, 13538.

[45]    X. Liang, H. Xia, J. Xiang, F. Wang, J. Ma, X. Zhou, H. Wang, X. Liu, Q. Zhu, H. Lin, J. Pan, M. Yuan, G. Li, H. Hu, *Advanced Science* **2024**, *11*.

[46]    S. Svanström, T. J. Jacobsson, G. Boschloo, E. M. J. Johansson, H. Rensmo, U. B. Cappel, *ACS Appl Mater Interfaces* **2020**, *12*, 7212.

[47]    G. Divitini, S. Cacovich, F. Matteocci, L. Cinà, A. Di Carlo, C. Ducati, *Nat Energy* **2016**, *1*, 15012.

[48]    B. Liu, R. R. Gutha, B. Kattel, M. Alamri, M. Gong, S. M. Sadeghi, W.-L. Chan, J. Z. Wu, *ACS Appl Mater Interfaces* **2019**, *11*, 32301.

[49]    Y. Dong, Y. Gu, Y. Zou, J. Song, L. Xu, J. Li, J. Xue, X. Li, H. Zeng, *Small* **2016**, *12*, 5622.

[50]    Y. Li, Z.-F. Shi, S. Li, L.-Z. Lei, H.-F. Ji, D. Wu, T.-T. Xu, Y.-T. Tian, X.-J. Li, *J Mater Chem C Mater* **2017**, *5*, 8355.

[51]　L. Li, S. Ye, J. Qu, F. Zhou, J. Song, G. Shen, *Small* **2021**, *17*.

[52]　R. K. Smith, P. A. Lewis, P. S. Weiss, *Prog Surf Sci* **2004**, *75*, 1.

[53]　S. Bang, N. T. Duong, J. Lee, Y. H. Cho, H. M. Oh, H. Kim, S. J. Yun, C. Park, M.-K. Kwon, J.-Y. Kim, J. Kim, M. S. Jeong, *Nano Lett* **2018**, *18*, 2316.

[54]　M. Makowski, W. Ye, D. Kowal, F. Maddalena, S. Mahato, Y. T. Amrillah, W. Zajac, M. E. Witkowski, K. J. Drozdowski, Nathaniel, C. Dang, J. Cybinska, W. Drozdowski, F. A. A. Nugroho, C. Dujardin, L. J. Wong, M. D. Birowosuto, *Advanced Materials* **2025**.